\documentstyle[12pt]{article}
         \makeatletter
         \def\thefigure{\@arabic\c@figure}\def\fps@figure{tbp}
         \def\ftype@figure{1}\def\ext@figure{lof}
         \def\fnum@figure{\protect\footnotesize Fig.\ \thefigure}
         \def\thetable{\@arabic\c@table}
         \def\fps@table{tbp}\def\ftype@table{2}\def\ext@table{lot}
         \def\fnum@table{\protect\footnotesize Table \thetable}
         \def\@listI{\leftmargin\leftmargini\parsep=0pt\itemsep=0pt}
         \def\thebibliography#1{\section{References}\vspace*{-10pt}\list
          {[\arabic{enumi}]}{\settowidth\labelwidth{[#1]}\leftmargin\labelwidth
          \advance\leftmargin\labelsep
          \usecounter{enumi}}
          \def\newblock{\hskip .11em plus .33em minus .07em}
          \sloppy\clubpenalty4000\widowpenalty4000
          \sfcode`\.=1000\relax}
         
         \def\@nomath#1{\ifmmode \fi}
         \def\mmycite{\@ifnextchar [{\@tempswatrue\@mmycitex}
             {\@tempswafalse\@mmycitex[]}}
         \def\@mmycitex[#1]#2{\if@filesw\immediate%
         \write\@auxout{\string\citation{#2}}\fi
           \def\@citea{}\@mmycite{\@for\@citeb:=#2\do
             {\@citea\def\@citea{,}\@ifundefined
                {b@\@citeb}{{\bf ?}\@warning
                {Citation `\@citeb' on page \thepage \space undefined}}%
         \hbox{\csname b@\@citeb\endcsname}}}{#1}}
         \def\@mmycite#1#2{{{\scriptsize#1}\if@tempswa , #2\fi}}
         \def\mycite#1{$^{\protect\mmycite{#1}}$}
         \def\@cite#1#2{{#1\if@tempswa , #2\fi}}
         \def\thesection {\arabic{section}}
         \def\section#1{\addtocounter{section}{1}\setcounter{subsection}{0}
              \bigskip\medskip{\noindent\bf\thesection.\ #1}
              \medskip}
         \def\thesubsection {\arabic{section}.\arabic{subsection}}
         \def\subsection#1{\addtocounter{subsection}{1}
              \medskip{\noindent\thesubsection.\ #1}
              \medskip}
         \makeatother
\begin{document}
\vspace*{0.3in}
\begin{center}
  {\bf DYNAMICAL CHAOS AND CRITICAL BEHAVIOR IN
  VLASOV SIMULATIONS OF 
   NUCLEAR MULTIFRAGMENTATION} \\

  \bigskip
  \bigskip
A. Atalmi, M. Baldo, G.\,F. Burgio and
{\underline {A. Rapisarda}} \\
  \bigskip

{\em CSFNSM,  
Dipartimento di Fisica Universit\'a  di Catania  \\
and I. N. F. N. Sezione di Catania \\
C.so Italia 57, I-95129 Catania, Italy}\\

  \bigskip
\end{center}
\smallskip
{\footnotesize
\centerline{ABSTRACT}
\begin{quotation}
\vspace{-0.10in}
\noindent
We discuss the presence of both dynamical chaos and signals of a
second--order phase transition in numerical  Vlasov simulations of  nuclear
multifragmentation. We find that chaoticity and criticality are strongly
related  and play a crucial role in the process of fragments formation.
This  connection is not limited to our model and seems a rather general
feature. 
\end{quotation}}


\section{ Introduction}

In the last decade the phenomenon of nuclear multifragmentation in 
heavy-ion reactions at intermediate energies and the possibility of 
observing a phase transition in nuclear matter has stimulated  many
experimental and theoretical  investigations
\mycite{purdue,ber84,gross,bondorf}.  Very recently the publication of
clear experimental evidence for a  phase transition \mycite{eos,gsi,tesi}
has confirmed the  first theoretical conjectures, reinforcing, on the other
hand, serious and puzzling questions. Can one really speak of a phase
transition in a finite and transient system? Are we facing a transition of
a  kind already known? What is the role of the long-range Coulomb forces?
Is it a  1st or 2nd order phase transition?  Can we distinguish between the
two in a finite system? Is statistical equilibrium reached? What is the
role of dynamical chaos? In the effort to answer to the above questions a
variety of dynamical and statistical models have been proposed in the 
past\mycite{gross,bondorf,bertsch,bur92,bau92}. Two main scenarios seem to
emerge from what we know at the moment.  At incident energies of 50-100
MeV/A  and for central collisions  collective flow and dynamical effects
seems to play a major role.  In this case, one expects that some memory of
the most important collective  excited modes should remain
\mycite{memo,fil}.  On the other hand, at higher energies  and for
peripheral collisions,  collective flow should become less important and 
thermal equilibrium is very likely reached\mycite{eos,gsi}.  In both cases,
it seems that  the hot system  formed in the first stage of the reaction
expands and  enters the spinodal region of  nuclear matter in the cooling
phase. It is at this moment that the process of fragment formation begins.
In the first scenario  the time scale  is very fast $\sim 50~ fm/c$, while,
in the other one, one has to wait  $\sim 150-200 ~fm/c$ before fragments
appears.  Prompt statistical models seem to describe quite well this second
scenario where some kind of thermalization is reached
\mycite{gross,bondorf}.  In the following we will consider only this kind
of multifragmentation. The break-up of the hot composite system  into
several big fragments with $Z \ge 3$ is however a very rapid process. At
the same time, it is also different from standard statistical phenomena
like  fission or compound nucleus formation, whose  typical time-scale  is
of the order of thousands fm/c.   It is therefore not clear why statistical
models\mycite{gross,bondorf} are able to explain the experimental data.  On
the other hand, their undoubted success indicates that the phase space
dominates the population of the final channels. In such a fast process, the
tendency of filling uniformly the phase space cannot occur in each
collisional  event. The system has not enough time to explore, during the
reaction,  the whole phase space, as can  happen for compound nucleus
collisions. It can be expected, therefore, that the reaction dynamics is
dominated by the phase space only when the physical quantities are averaged
over large  sets of events. This assumption implies that the nuclear
dynamics in the  multifragmentation regime is irregular or chaotic enough
to produce, at least approximately, a uniform "a priori" probability  to
populate each region of the available phase space. In particular, the
formation of the final fragments must follow  an irregular dynamics. This
conjecture can be also inferred by the large  event-to-event fluctuations
observed experimentally on the charge and  mass distributions. In this
respect the hypothesis  that multifragmentation could be reminiscent of a
phase transition finds a natural 
explanation\mycite{campi,bauer,plosz,jaq,vito}. 
The crucial role of deterministic  chaos in multifragmentation was found in
Vlasov-Nordheim simulations in refs.\mycite{bbr1,bbr2,bbr3} and it was
recently confirmed by  Classical Molecular Dynamics 
calculations\mycite{aldo} in connection with a critical behavior. In the
following, we discuss nuclear multifragmentation  along the same line of
refs.\mycite{bbr1,bbr2} presenting the latest results. The model was first
proposed in ref.\mycite{bur92} and it represents a schematic but precise
and instructive  guideline for understanding the dynamics of fragment
formation \mycite{bbr2}. After a short description of the model in section
2, we report in section 3  recent calculations\mycite{bbr3}, which  show
how rapidly the  system reaches  statistical equilibrium by means of
deterministic chaos.  In particular scatter plots of the excited modes are
studied as a function of time and fractal dimensions are calculated. In
section 4,  we show for the first time that, within this mean-field
approach, a critical behavior can be also obtained, if an event-to-event
analysis is performed. Conclusions are drawn in section 5, where  we
conjecture that, the strong relation between  phase transitions and
chaoticity found in our simulations is a rather  general feature, as
confirmed by several recent  studies\mycite{aldo,nayak,yama}.

\section{ The model}

The Vlasov-Nordheim  equation was solved  numerically in a two-dimensional
lattice \mycite{bbr1,bbr2,bbr3} using the same code of ref.\mycite{bur92}.
The collision integral was actually considered only in a few simulations
\mycite{bbr2} not reported in this paper. Actually it seems that two-body
collisions change only slightly  the features here discussed. In the
following we will refer only to Vlasov  simulations. We studied a fermion
gas situated on a large torus  with periodic boundary conditions, and its
size was kept constant during the evolution. The torus sidelengths are
equal to $L_x = 51~fm$ and $L_y = 15~fm$. We divided the single particle
phase space into several  small cells. We employed in momentum space 51x51
small cells of size $\Delta p_x = \Delta p_y = 40~ MeV/c$, while in
coordinate space $\Delta x = 0.3333~fm$ and $\Delta y = 15~fm$, $i.e.$ we
considered  only one big cell on the $y$-direction. The initial local
momentum distribution was assumed to be the one of a  Fermi gas at a fixed
temperature $T = 3~MeV$. We employed  a local Skyrme interaction which
generates a mean field  $U[\rho] = t_0~(\rho / \rho_0) + t_3~(\rho /
\rho_0)^2$. The density $\rho$ was folded along the $x$-direction with a
gaussian $e^{-x^2/\mu^2}$  with $\mu = 0.61 fm$, in order to give a finite
range to the interaction. The parameters of the force $t_0$ and $t_3$ were
chosen in order to reproduce correctly the binding energy of nuclear matter
at zero temperature, and this gives $t_0=-100.3~MeV$ and $t_3 =48~MeV$. 
The resulting EOS gives a saturation density in two dimensions equal to
$\rho_0 =0.55~fm^{-2}$ which corresponds to the usual three-dimensional 
Fermi momentum equal to $P_F=260~MeV/c$. The step adopted for the time
evolution  was equal to $\Delta t=0.5~fm/c$. For more details concerning
the model see  refs.\mycite{bur92,bbr1,bbr2}.

\section{ Multifragmentation and Deterministic Chaos}

In order to investigate the regular or irregular behavior of the 
mean-field dynamics  with respect to multifragmentation, we studied the
response of the system   to small initial perturbations inside the spinodal
region of nuclear matter. In our simulations we neglect the initial part of
the  process which  drives the system inside the spinodal zone. Therefore
the most  realistic way to perturb our system is obtained  by imposing a
uniform and small  white noise on the average density profile. This random
initialization will mimic  the initial uncertainties.  

In fig.1 we show the time evolution of the density profiles for   two
random-initialized events started at half the saturation density.  The
initial perturbation was $1\%$ of the average density. Notice the different
scale used in the upper panels of the figure in order to magnify  the
initial noise.  Though this perturbation is extremely small at t=0  and the
two initializations are very close in phase space, fluctuations are
amplified and distorted during the time evolution. The two simulations
evolve following different deterministic paths in a complete unpredictable
way. As a further  evidence of this behavior, the power spectra
corresponding to the  density profiles are shown in fig.2. The broad range
of modes $n_k$ initially excited ($n_k=k L_x/2 \pi$)  evolve differently in
each run. No particular mode is privileged by the dynamics apart from a
natural  cut-off of the highest modes due the finite range of the
interaction.  For more details see refs. \mycite{bbr1,bbr2,bbr3}.

\begin{figure}
 \begin{center}
\vspace*{2.5in}
 \end{center}
\caption[]{\footnotesize
Time evolution of two density profiles  started at half the saturation
density with a very small random initialization.  The strength of this
white noise was only $1\%$ of the average initial density. Please notice
that in the upper panels for t=0  the scale is magnified.
}
\end{figure}
\begin{figure}
 \begin{center}
\vspace*{2.5in}
 \end{center}
\caption[]{\footnotesize
Power spectra corresponding to the density profile evolution
of the previous figure.
}
\end{figure}

The above arguments give already a first qualitative idea of the
sensitivity to the initial conditions and the irregularity of the 
following evolution which are  typical features of deterministic chaos.

A way to quantify chaotic behavior is by calculating the  largest Lyapunov
exponent $\lambda$,  i.e. the average rate of divergency between  two close
trajectories in phase space. This calculation was done in refs.
\mycite{bbr1,bbr2} where a value ranging from 0.028 to 0.1 c/fm was
extracted according to the initial density considered inside the  spinodal
region.  Outside this zone, on the other hand, $\lambda=0$ and the dynamics
is regular.  Actually, due to the unbound dynamics and its limited time
scale, the finite and positive value obtained for the largest Lyapunov
exponent is not a unique signature of chaoticity.  For a critical review of
this quantity see ref.\mycite{bbr3}. However, it should be noticed that,
the  sensitive dependence on the initial conditions shown in   figs.1,2 and
the positive Lyapunov exponent found, leave no room to ambiguities of any
sort. 

The time scale related to the chaotic behavior, $\tau=1/\lambda$, is of the
order of $\sim$10-30 fm/c. If one adds to this time the  transient time
interval of $\sim$ 20 fm/c needed by the largest exponent to prevail on the
others,  then  the mean-field evolution  becomes irregular and
unpredictable after $\sim$40-50 fm/c. This is in perfect agreement with
what found in ref.\mycite{fil}, where (see fig.3 of ref.\mycite{fil})
linear response reproduces the numerical simulations only up to $\sim50
~fm/c$. 

A further and impressive way to show the onset of chaoticity can be
obtained by plotting the final amplitudes of the excited modes as a
function of the  initial ones. This technique  has been widely used for
chaotic scattering \mycite{chaos,rap91} and it was adopted in
ref.\mycite{bbr2} for two coupled harmonic oscillators. In fig.3 we show
these  scatter plots as a function of time  for 500 random-initialized
events at an average density  $\rho/\rho_0=0.5$. Only the modes $n_k=5,14$ 
are shown\mycite{bbr3} . Similar plots were done independently in
ref.\mycite{jac}. The figure illustrates very clearly that  after an
initial linear and regular evolution, whose time scale depends on the mode
considered, the onset of chaos  leads to the appearance of wild
fluctuations. Please notice that the numerical algorithm is under control
and the total energy is conserved within $1\%$. The fluctuations are  due
only to the non-integrability of the Hamiltonian which  produces this
irregular evolution. In other words, the fluctuations are an intrinsic
feature of the  deterministic equations considered and not a spurious
numerical noise. It is important to stress that   the  time of fragment
formation $\sim$100-150 fm/c (see fig.1) is much longer  than the time for
chaos onset, therefore multifragmentation is strongly affected by this
irregular evolution.

\begin{figure}
 \begin{center}
\vspace*{2.5in}
 \end{center}
\caption[]{\footnotesize
Scatter plots of the initial and final amplitudes 
for the modes $n_k=5$ and $n_k=14$. The evolution of
500 simulations of the same kind shown in fig.1 is considered at 
different time scales. 
See text for more details.}
\end{figure}
\begin{figure}
 \begin{center}
\vspace*{2.5in}
 \end{center}
\caption[]{\footnotesize 
Artificial random distribution in one (a) and two (c) dimensions. A set of
500 points is considered. The logarithm of the corresponding correlation
integral $C(r)$   is plotted vs. $ln~r$ in the lower panels (b) and (d) 
(circles).  The fractal dimension $D_2$, given by the slope of the fit
(full line), coincides (as it must be)  with the euclidean dimension.
}
\end{figure}

In order to quantify the dispersion  of the points plotted in fig.3 one can
calculate the fractal  correlation dimension $D_2$ \mycite{chaos} using the 
Grassberger and Procaccia correlation integral\mycite{grass}. The latter is
defined as 
\begin{equation}
C(r) = {1\over M^2} \sum_{i,j}^M \Theta ( r - |{\bf z}_i - {\bf z}_j|) ~~,
\end{equation}
$\Theta $ being the Heaviside step function, ${\bf z}_i$ a vector whose 
two components ($x_i, y_i$) are the initial  and final amplitudes of the
modes, $M$ the total number of points and  $r$ a sampling interval. For
small $r$  one gets $C(r)\sim r^{D_2}$. $D_2$ is the fractal correlation
dimension. It has been used in nonlinear dynamics  to study how often  each
part of a  strange attractor is visited. In general it gives information on
how filled is the phase space. For more details see
ref.\mycite{chaos,grass,pal}.

\begin{figure}
 \begin{center}
\vspace*{2.5in}
 \end{center}
\caption[]{\footnotesize
The behavior of $ln~C(r)$ vs. $ln~r$ is plotted (circles) for the modes
$n_k=5$ (a) and $n_k=14$ (b)  at t=120 fm/c.  The full line is the linear
fit, whose slope gives the fractal dimension $D_2$ reported. See text for
more details.}
\end{figure}


In fig.4 (b)(d) we display as an example the scaling behavior of $ln~C(r)$
versus $ln~r$ (open circles) for  two sets of points distributed at random
on a line (a)  and on a plane (c). As expected one gets, from the slopes of
the linear fits (full line), a value of $D_2$ equal to the euclidean
dimension  within the errors. The latter are  due to the limited number of
points considered. In this case, in order to test the accuracy of the
algorithm  only 500 points  were taken into account. So the method is
reliable to calculate $D_2$ in the case of our Vlasov scatter plots.

In fig.5 we show the scaling behavior  obtained for the scatter plots of
fig.3  at t=120 fm/c. The full line is the linear  fit of the numerical
simulation (open circles), which enables to extract the  value
$D_2=1.96\pm0.05$ and $D_2=1.94\pm0.06$ respectively.  Finite size effects
limit this scaling which in any case extends to several decades
\mycite{bbr3}. The behavior of the Vlasov calculations is very similar  to
that obtained for the 2D random distribution  of fig.4 (c,d). This means
that at t=120 fm/c for both modes we have complete randomness.

The time evolution of $D_2$ from 1  to 2 is clearly evident in  fig.6.  The
calculations refers to the scatter plots of fig.3. In general, while the
onset of deterministic  chaos starts around 40-50 fm/c for the lowest
modes, the dynamics is more regular for the highest ones. However after
$\sim$80 fm/c the  evolution tends to a complete random distribution. 

This result is true for all the modes excited and leaves no room for
regularity of any sort. In other words there is a clear tendency towards 
equilibration and complete filling of phase-space. This is just what all
the statistical models \mycite{gross,bondorf} assume as a starting point.
Hence these schematic  numerical simulations demonstrate that, even within
a mean-field  description, one can reach statistical equilibration in a
very short  time scale if the system enters the spinodal region where
chaoticity is at work.


\begin{figure}
 \begin{center}
\vspace*{2.5in}
 \end{center}
\caption[]{\footnotesize
Correlation dimension $D_2$ as a function of time 
for the irregular evolution of the modes $n_k=5$ and $n_k=14$
displayed in fig.3.}

\end{figure}

\section{ Critical behavior}

In general sharp second--order  phase transitions can be observed only in
macroscopic systems \mycite{stan,fisher}, however also in small finite
systems one can find  clear signals of critical behavior
\mycite{campi,bauer,elliot,mueller}. In finite systems the  singularities
are smoothed, but one can still extract critical exponents  which are very
close to those of the infinite system.  By using
percolation\mycite{campi,bauer},  classical molecular dynamics
\mycite{vito} and statistical  models \mycite{jaq} it has been claimed that
nuclear matter can show signals of a second--order phase transition.  Some
preliminary indication of criticality was observed many years
ago\mycite{purdue,wad,campi}, but it was only very recently  that critical
exponents  were extracted from exclusive experimental nuclear
data\mycite{eos,tesi}.

In general, one studies cluster size distributions by considering the
conditional moments 
\begin{equation}
M_k = \sum s^k n(s,\epsilon) ~~,
\end{equation}
where $n(s,\epsilon)$ is the number of fragments of size $s$ and $\epsilon$
is a variable which indicates the distance  from the critical point. For
thermal transitions $\epsilon=T_c-T$, while  for percolation
$\epsilon=p-p_c $, $T_c$ and $p_c$ being respectively  the critical
temperature and the threshold probability.  The summation runs over all
fragments except the heaviest one.  Near the critical point one gets the
scaling behavior
\begin{equation}
n(s,\epsilon) \sim  s^{-\tau} f(\epsilon s^\sigma) ~~~,
\end{equation}
$\tau$ and $\sigma$ being two critical exponents.
At the critical point  $\epsilon = 0$  
and   $f(0) = 1 $, therefore the  
cluster size distribution shows a power law. Moreover the k-moments
obey the scaling relation
\begin{equation}
M_k \sim |~\epsilon~|^{-(1+k-\tau)/\sigma} ~~~.
\end{equation}
In particular for the first three moments one gets
\begin{equation}
M_0 \sim  |~\epsilon~|^{2-\alpha} ~~~,
M_1 \sim  (~\epsilon~)^{\beta} ~~~,
M_2 \sim  |~\epsilon~|^{-\gamma} ~~~,
\end{equation}
where $\alpha$, $\beta$ and $\gamma$ are characteristic critical exponents.

Critical exponents are not all independent and satisfy the relation
\begin{equation}
 \gamma + 2\beta = 2 -\alpha = {(\tau-1)\over{\sigma}} ~~~.
\end{equation}
In the infinite system, at the critical point, the moments with $k\ge2$
 diverge and also the correlation length $\xi$ diverges as 
\begin{equation}
\xi  \sim  |~\epsilon~|^{-\nu} ~~~,
\end{equation}
where $\nu$ is another critical exponent related to those above discussed
\mycite{stan,fisher}.
Therefore, at the critical point, fluctuations become very large and
involve the whole system  even if the microscopic forces have a very small
range.  The details of this microscopic interactions are not important any
longer and  one observes universal features.  There are several classes of
universality - percolation and liquid-gas  are  two examples. Each class
has   different critical exponents. We report in table 1 the value of some 
exponents for  percolation (2D and 3D) and liquid-gas phase transition. 
For a more detailed discussion on critical behavior see
ref.\mycite{stan,fisher,stauf}.

In analyzing nuclear data  it is difficult to identify a model-independent
critical control parameter. In general one cannot define an average cluster
size distribution   $n(s,\epsilon)$. In ref.\mycite{campi}  it was shown
that,  in order to avoid this difficulty,  an event-to-event analysis 
could be performed to investigate  critical behavior. In particular one can
define the conditional moments for each single event  as
\begin{equation}
M_k^j = \sum s^k m^j(s) ~~~,
\end{equation}
where $m^j(s)$ is the number of fragments of size $s$ in the event $j$. The
summation runs over all fragments, excluding the heaviest one  produced in
the event.

It was just by using such moments that Campi\mycite{campi} found in the
inclusive nuclear  data by Waddington and Freier\mycite{wad} clear
signatures of criticality. Along the same lines the analysis was performed 
for statistical\mycite{jaq} and dynamical models\mycite{vito}. 

In what follows, by means of the conditional moments defined in eq.(8), we
will show for the first time  that it is possible to obtain signals of
criticality also  within Vlasov simulations of nuclear multifragmentation. 
This fact is important since, in our case,  criticality is induced by
chaotic dynamics.


\begin{figure}
 \begin{center}
\vspace*{2.5in}
 \end{center}
\caption[]{\footnotesize
The logarithm of the largest cluster size $P$ is  plotted versus the
logarithm of the second moment  $M_2$ for an ensemble of 954 Vlasov
simulations. Each event was generated  with  a random initialization as in
fig.1 and  inside the spinodal region. Different starting average densities 
- in the range $\rho/\rho_0=0.4-0.6$ - were  considered. See text for
further details. 
}
\end{figure}

In ref.\mycite{bbr2} it was already discussed the fact that  in our Vlasov
numerical simulations of nuclear multifragmentation deterministic chaos
produces broad cluster size distributions and strong event-to-event
fluctuations so typical of multifragmentation data. In particular a power
law behavior was found with an exponent $\tau\sim2.17\pm0.3$. In that  case
the amount of events considered was very small, $\sim$100, and one could
not go beyond that. In the following, investigating  a larger  ensemble
consisting of 954 random-initialized multifragmentation events started
inside the spinodal region, we will present some preliminary  results which
show unambiguous signals of critical behavior.  More precisely, as
previously discussed,  we  perturbed  our system with a small white noise
inside the spinodal region. Several  initial average densities ranging from
0.4 to 0.6 $\rho/\rho_0$ were taken into account. Then  we followed  the
Vlasov time evolution as in fig.1 until well defined clusters were formed.
The time scale to get final fragments  ranges from $\sim$300 fm/c for
$\rho/\rho_0=0.6$ to  $\sim$80 fm/c for $\rho/\rho_0=0.4$.  The size of the
fragments was taken along the x-direction, considering all those
neighbouring cells whose density was bigger than the freeze-out one. We
considered as freeze-out density the value $\rho_{freeze-out}=0.1~
fm^{-2}$, which corresponds  to $\sim 20 \% ~\rho_0$. We have checked that
small variations of the freeze-out density do not produce significant
changes\mycite{bbr2}.


\begin{figure}
 \begin{center}
\vspace*{2.5in}
 \end{center}
\caption[]{\footnotesize
We plot $ln~P$ versus  $ln~<M_2>$ for the same  ensemble of the previous
figure (open circles).  The full lines are linear fits of the two branches.
The slopes of the  fits give the ratio of the critical exponents
$\beta/\gamma$. More precisely the upper (subcritical) one gives
$-\beta/\gamma=-0.27\pm0.1$, while the lower (overcritical) one gives $1 +
\beta/\gamma=1.46\pm0.1$. 
}
\end{figure}

In fig.7 we plot the logarithm of the size of largest fragment $P$ versus
the  logarithm of the second moment $M_2$. The figure shows a typical
critical triangular pattern with a  subcritical upper branch and an
overcritical lower one. 'Evaporation-like' events in the  upper branch are
obtained for initial densities  $\rho/\rho_0\sim0.6$, whereas
'vaporization-like' events are found for  $\rho/\rho_0\sim0.4$. Critical
events are mostly in the range   $ \rho/\rho_0\sim0.55-0.5$.  In general
large fluctuations exist from event to event.  The two branches meet at the
critical point.  Similar plots were also obtained in refs.\mycite{campi}
for  cubic bond percolation model with a number of sites $A=6^3$ and  the
gold multifragmentation emulsion data by Waddington and 
Freier\mycite{wad}.
\begin {table}
\begin{center}
\begin{tabular}{|c|c|c|c|c|} 
\hline\hline
$exponent$                     &  $\beta$  &  $\gamma$  & 
$\beta/\gamma$  &  $\tau$   \\ 
\hline
$2D ~Percolation$\mycite{stauf}    &    0.14   &    2.4    
&   0.058          &  2.0    \\ 
\hline
$3D ~Percolation$\mycite{stauf}    &    0.41   &    1.8    
&   0.23           &  2.18    \\ 
\hline
$Liquid-gas$\mycite{stan}      &    0.33   &   1.23   
 &   0.27           &  2.21    \\ 
\hline
$Emulsion~ data$\mycite{campi} &    -      &   -       
&   0.2$\pm$0.1    &  2.17$\pm$0.1       \\ 
\hline
$EOS~ data$ \mycite{eos}       & 0.29$\pm$0.02  &   
1.4$\pm$0.1  &  0.21$\pm$0.1  &  2.14$\pm$0.06 \\ 
\hline
$Vlasov$                       &    -   &   -    &   
0.37$\pm$0.2      &  2.27$\pm$0.2    \\ 
\hline \hline
\end{tabular}   
\caption{Comparison of Vlasov critical exponents with those 
of percolation (2D and 3D), liquid-gas and experimental nuclear data. }

\label {tab1}
\end{center}
\end{table}  

The indication given by fig.7 is important but qualitative. One should
extract critical exponents to have a quantitative information.  Following
the procedure adopted in ref.\mycite{jaq} we took the average  of $M_2$ at
a fixed size of the largest cluster P.  In this way one can try to extract
the slopes of the two branches  which should give the ratio between the
critical exponents  $\beta$ and $\gamma$. In fact the lower branch should
have a slope given by  $1 + \beta/\gamma$, while the upper branch
$-\beta/\gamma$\mycite{campi,jaq}.  In fig.8 we plot $ln~P$ as a function
of the logarithm of the  average $<M_2> $ (open circles).  In the figure
are shown also the linear fits (full lines)  of the two branches and the
values  of the slopes obtained.  The statistics of the upper branch is 
poorer than  that of the lower branch - as can be seen in fig.7. For this
reason the  first point of the upper branch was excluded from the fit. From
the two slopes one gets consistent values within the errors, i.e.
$-\beta/\gamma=-0.27\pm0.1$ and $1 + \beta/\gamma=1.46\pm0.1$  Taking an
average between the two values  one finds $\beta/\gamma = 0.37 \pm 0.2$. 
The ratio between $\beta$ and $\gamma$ can be used to extract the value of 
$\tau$. In fact from eqs.(4-6) one obtains the relation
\begin{equation}
 \tau = { {2 + 3 \beta/\gamma} \over {1 + \beta/\gamma} } ~~~ .
\end{equation}
From the above equation and our estimate of $\beta/\gamma = 0.37 \pm 0.2$
we get $\tau=2.27 \pm 0.2$, which is consistent to what previously 
obtained\mycite{bbr2}. 

It is important to stress at this point, that we start our system inside
the spinodal region, thus it is not clear at the moment what kind of  phase
transition the system is reminiscent of. We find signals of criticality in
the sense that the instabilities of the spinodal zone, driven by
deterministic  chaos, produce inclusive cluster size distributions of 
different shapes, as in percolation or liquid-gas phase transitions,
according to the initial average density adopted.  At variance with
percolation, which is a geometric model, we have a nuclear-like dynamics. 
It is not clear what are the links, if any,  between this behavior and the
standard liquid-gas phase transition.  The system in fact should not be
close to the critical point\mycite{bbr2}.  A deeper understanding of this
kind of criticality is beyond the aim of the present  paper and is left for
the future\mycite{future}.

In table 1 we report the value of the critical exponents $\beta$, $\gamma$
and $\tau$ obtained in our calculations together with those of  percolation
(in 2D and 3D) and the liquid-gas case. The values of the
Waddington\mycite{campi} and EOS data\mycite{eos} are also shown for
comparison.  In general the values we get from our Vlasov simulations are
not so  different from the 3D percolation and the liquid-gas phase 
transition. However, before  drawing definite conclusions some comments
must be done. i) The numerical calculation of critical exponents  is a very
delicate problem and our present analysis is rather crude.  ii) The
statistics we used is rather poor, expecially for subcritical events.  iii)
We have no idea of the absolute values of $\beta$ and $\gamma$.  iv) Our
simulations are in 2 dimensions and the cluster sizes were considered only
along the x-coordinate. 

On the other hand, it is very remarkable that  a schematic model like ours
is able to give such a critical behavior. Please notice also the close
similarity of our values with those extracted from the available
experimental data. This fact deserves a further comment since we neglected
the coulomb force. In fact, according  to statistical
calculations\mycite{jaq},   the coulomb force can strongly influence the
critical exponents. However no evidence of this influence seems to be
present in the available experimental data. A further puzzling question is
the understanding of the link between  this critical behavior and  
liquid-gas transition. 

As a general statement, we can say that the above discussed results  though
very impressing and stimulating  pose many troubling  questions and leave
several open problems.  A more severe analysis in order  to clarify the
situation is in progress\mycite{future}.

\section{Conclusions}

In this paper we concisely reviewed recent  numerical simulations of
nuclear multifragmentation. The growth of density fluctuations, in  the
spinodal region of nuclear matter EOS, has been advocated as the mechanism
of fragment formation. We studied the dynamics of such a  process by
solving the Vlasov equation numerically on a lattice. The nuclear system in
the spinodal region was schematized by a  two--dimensional fluid confined
inside a fixed torus. The initially  homogeneous system was perturbed by
adding a small and random density fluctuation. This was done  in order to
simulate the initial  uncertainties produced by the missing initial
dynamics. The time evolution was then followed up to the time of fragment
formation, when several well separated density humps appear.  The study of
scatter plots of the excited modes confirmed the chaotic character of the
dynamics found already in refs.\mycite{bbr1,bbr2}. In the present
contribution it was shown that chaos is strong enough to fill uniformly the
available phase space. This fact was quantified  by the calculation of the
fractal correlation dimension $D_2$.  Chaos is fully developed  in a short
and limited time interval, of the order of 80-120 fm/c\mycite{bbr3}. In
this respect such a chaotic dynamics is a transient phenomenon  similar to
the case of chaotic scattering\mycite{rap91}. 

These  results give  a strong theoretical support to the statistical
models\mycite{gross,bondorf} successfully used in analyzing most of the
available nuclear  multifragmentation data.

Finally, using an ensemble of  954 Vlasov simulations  started inside the
spinodal region, an event-to-event analysis of the cluster size
distributions  was  performed along the same lines of
refs.\mycite{campi,jaq}. Unambiguous signals of  critical behavior were
found also in our case.  The extracted ratio of the  critical exponents
$\beta/\gamma$ and of $\tau$ are in agreement with those  of liquid-gas, 3D
percolation and the available experimental  data\mycite{campi,wad,eos}.  At
the moment both the analysis and the model are too schematic to 
discriminate among different universality classes or to  draw final
conclusions. Here we do not want to stress too much the  values of the
critical exponents obtained, but it seems very important that both 
chaoticity and criticality are found in the process of multifragmentation. 
Very recently a similar coexistence was  found  also in other
models\mycite{aldo,nayak,yama}. This fact confirms the  general character
of this feature, which could probably be explained by the enhancement of
phase-space  volume and entropy during the phase transition. Chaoticity is
probably another  indication of phase transition. Such an appealing 
general perspective is at the moment only a sound conjecture which  needs
further investigation.

\noindent

\section{Acknowledgments}

We would like to thank 
M. Belkacem, A. Bonasera and V. Latora for several stimulating discussions.

\end{document}